\documentclass[preprint,5p,times,twocolumn]{elsarticle}
\usepackage{amsmath,amssymb}
\usepackage{type1cm}
\usepackage{mathrsfs}
\usepackage{url}
\usepackage{graphicx}
\usepackage{color}
\usepackage{booktabs}
\usepackage[switch]{lineno}

\allowdisplaybreaks[1]
\newcommand\HL[1]{{\color{black}#1}}
\newcommand\HLL[1]{{\color{black}#1}}

\def\vec#1{{\rm  \bf #1}}
\journal{Journal}

\begin{document}
\begin{frontmatter}

\title{Entropy stability analysis of smoothed dissipative particle dynamics}

\author[RCAST]{Satori Tsuzuki}
\ead{tsuzuki.satori@mail.u-tokyo.ac.jp}
\address[RCAST]{Research Center for Advanced Science and Technology, The University of Tokyo, \\4-6-1, Komaba, Meguro-ku, Tokyo 153-8904, Japan\\ 
}

\begin{abstract}
This article presents an entropy stability analysis of smoothed dissipative particle dynamics (SDPD) to review the validity of particle discretization of entropy equations. First, \HL{we consider the simplest SDPD system: a simulation of incompressible flows using an explicit time integration scheme, assuming a quasi-static scenario with constant volume, constant number of particles, and infinitesimal time shift.} Next, we derive a form of entropy from the discretized entropy equation of SDPD by integrating it with respect to time. We then examine the properties of a two-particle system \HL{for a constant temperature gradient}. \HL{Interestingly, our theoretical analysis suggests that there exist eight different types of entropy stability conditions, which depend on the types of kernel functions. It is found that the Lucy kernel, poly6 kernel, and spiky kernel produce the same types of entropy stability conditions, whereas the spline kernel produces different types of entropy stability conditions.}
Our results contribute to a deeper understanding of \HL{particle discretization.}
\end{abstract}
\end{frontmatter}

\section{Introduction}
The smoothed dissipative particle dynamics (SDPD), which was proposed by Espa{\~n}ol~(2003)~\cite{PhysRevE.67.026705}, has attracted the attention of many physicists and engineers for over a decade. The SDPD is appropriate for simulations of mesoscale flows in which thermal fluctuations are non-negligible, \HL{and it} has been acknowledged in many complex flow problems ({\it e.g.,} cellular blood flows~\cite{ALIZADEHRAD2018303, PhysRevE.95.063314} and suspension flows including polymer molecules~\cite{PhysRevE.77.066703}). By introducing conservation of angular momentum, the \HL{accuracy of SDPD simulations of} incompressible flows has been \HL{greatly improved}~\cite{doi:10.1063/1.2359741, MULLER2015301}. 
Thus, the SDPD exhibits great potential for solving many types of thermal flow problems that exist around us.

As described in the literature~\cite{PhysRevE.67.026705}, the governing equations of hydrodynamics with thermodynamic consistency are 
\begin{eqnarray}
\frac{d\rho}{dt} &=& -\rho\vec{\nabla}\cdot\vec{v}, \label{eq:gvn:1}\\
\rho\frac{d\vec{v}}{dt} &=& -\nabla P+\eta \nabla^2 \vec{v}+\Biggl(\zeta +\frac{\eta}{3}\Biggr) \nabla\nabla\cdot\vec{v}, \label{eq:gvn:2}\\
T\rho\frac{ds}{dt} &=& \phi+\kappa \nabla^2 T. \label{eq:bachelor:gvneq}
\end{eqnarray}
Here, $\rho$,~$\kappa$,~$\eta$, and~$\zeta$ are density, thermal conductivity, sheer viscosity, and bulk viscosity, respectively. 
The set of Eq.~(\ref{eq:gvn:1}) and Eq.~(\ref{eq:gvn:2}) represents the Navier--Stokes equations, and Eq.~(\ref{eq:bachelor:gvneq}) represents the relationship among the entropy $s$, viscous heating field $\phi$, and temperature field $T$. Hereinafter, we refer to Eq.~(\ref{eq:bachelor:gvneq}) as the `Batchelor's equation', as in~\cite{batchelor1967introduction}.

Because the SDPD is a kind of Lagrangian particle method, each of Eq.~(\ref{eq:gvn:1}) to Eq.~(\ref{eq:bachelor:gvneq}) is discretized using particles in a similar manner to smoothed particle hydrodynamics (SPH)~\cite{gingold1977smoothed, morris1997modeling}. 
According to Eq.~(32) in ~\cite{PhysRevE.67.026705}, the discretized expression for the rate of change of the total entropy $S$ is represented as
\begin{eqnarray}
\frac{dS}{dt} &=& \sum_{i}\frac{\phi_{i}}{T_{i}} + \kappa \sum_{ij} \frac{F_{ij}}{d_{i}d_{j}T_{i}T_{j}} {T_{ij}}^2. \label{eq:batcheler1967} 
\end{eqnarray}
\HL{Here, $T_{i}$ and $T_{j}$ are the temperature of the $i$th and $j$th particles, respectively.
	The viscous heating field $\phi_{i}$, parameter $d_{i}$, relative temperature $T_{ij}$, and relative function $F_{ij}$ are given as}
\begin{eqnarray}
	\phi_{i} &=& \Biggl( \frac{5\eta}{6}-\frac{\zeta}{2}\Biggr)\sum_{j}\frac{F_{ij}}{d_{i}d_{j}}\vec{v}_{ij}^{2} \nonumber \\ 
	         &+& \frac{5}{2}\Biggl( \zeta + \frac{\eta}{3}\Biggr) \sum_{j}\frac{F_{ij}}{d_{i}d_{j}} {({\rm \bf e}_{ij} \cdot \vec{v}_{ij})}^{2},  \label{eq:phibreak} \\
d_{i} &=& \sum_{j} W(|\vec{r}_{ij}|),  \label{eq:dijbreak} \\
T_{ij} &=& T_{i}-T_{j},  \label{eq:Tijbreak} \\
F_{ij} &=& F(\left|\vec{r}_{ij}\right|).  \label{eq:Fijbreak}
\end{eqnarray}
\HL{where the relative vector $\vec{r}_{ij}$, unit vector $\vec{e}_{ij}$, relative velocity $\vec{v}_{ij}$, kernel function $W$, and its gradient function $F$ are given as}
\begin{eqnarray}
\vec{r}_{ij} &=& \vec{r}_{i}-\vec{r}_{j},  \\
\vec{e}_{ij} &=& \frac{ \vec{r}_{ij}}{|\vec{r}_{ij}|},  \label{eq:eijbreak}\\
\vec{v}_{ij} &=& \vec{v}_{i}-\vec{v}_{j},  \label{eq:vijbreak}\\
W(r) &=& \frac{105}{16 \pi h^3}\Biggl(1+3\frac{r}{h}\Biggr)\Biggl(1-\frac{r}{h}\Biggr)^3, \label{eq:kernel}\\
F(r) &:=& \frac{1}{\vec{r}} \nabla W(r) \nonumber \\
  	 &=& \frac{315}{4\pi h^5}\Biggl(1-\frac{r}{h}\Biggr)^2.  \label{eq:F}
\end{eqnarray}
Additionally, the parameter $h$~$(h > 0)$ is the kernel radius that determines the interaction range between particles.
\HL{The function $W$ (Lucy kernel function~\cite{lucy1977numerical}) in Eq.~(\ref{eq:kernel}) and the function $F$ in Eq.~(\HLL{\ref{eq:F}}) are the reposts of Eq.~(9) and Eq.~(10) in~\cite{PhysRevE.67.026705}, which presents the original SDPD.}

Equation~(\ref{eq:batcheler1967}) represents the introduction of a weighted calculation depending on the positions of particles using the function $F$. \HL{Although computational physicists have discussed the validity of applying a particle discretization scheme to compute physical quantities in the case of density calculation of SPH~\cite{gingold1977smoothed, morris1997modeling} or that of moving particle semi-implicit~(MPS)~\cite{koshizuka1996moving, koshizuka2018moving},} \HL{the validity of applying a particle discretization scheme to the entropy calculation in Eq.~(\ref{eq:batcheler1967}) has not been corroborated in terms of theoretical thermodynamics, even though the mathematical derivation of Eq.~(\ref{eq:batcheler1967}) and its expression in the GENERIC framework~\cite{PhysRevE.56.6620, PhysRevE.56.6633, PhysRevE.57.1416} were discussed in the original research~\cite{PhysRevE.67.026705}}.

The primary purpose of this article is to discuss the thermodynamic validity of Eq.~(\ref{eq:batcheler1967}) in SDPD.
 \HL{Our strategies are as follows. First, we consider a quasi-static scenario with constant volume, constant number of particles, and infinitesimal time shift, assuming the moment of an explicit SDPD simulation of incompressible flows. Next, we derive a form of entropy from the discretized entropy equation of SDPD by integrating it with respect to time. We then examine whether the SDPD system exhibits physically reasonable behaviours} by performing thermodynamic entropy analysis.

The remainder of this article is structured as follows. Section 2 derives the form of entropy from Eq.~(\ref{eq:batcheler1967}). In Section 3, we examine the characteristics of a two-particle system of SDPD by examining entropy stability conditions. Finally, Section 4 summarizes our results and concludes the article.

\section{Derivation of entropy}\label{sec:DERIVATION}
Let us consider the case that time $t$ varies from $t_0$ to $\tau$, and temperature $T_{i}$ of the $i$th particle~($i=1,2,\dots, N$) varies from $T_{i}^{0}$ to $T_{i}^{\tau}$ \HL{in Kelvin}.
Let the time $\tau$ {be} between $t_{0}$ and $t_{0} + \Delta t$.
\HL{In this paper, we focus on the simplest SDPD system, which is an explicit simulation of incompressible flows using a forward-Euler time integration scheme~\cite{corless2013graduate}.} 
\HL{Under this premise}, the \HLL{left}-hand side of Eq.~(\ref{eq:batcheler1967}) is approximated by the ratio of infinitesimal entropy $\Delta S$ to infinitesimal time $\Delta t$ 
$(dS/dt\ \approx \Delta S/\Delta t)$ in the simulation space. The positions \HL{and velocities} of all particles are fixed during the interval of $\Delta t$. Also, the values of $d_{i}$, $d_{j}$, and $F_{ij}$ become constant in this interval. \HL{Accordingly}, the viscous heating field $\phi_{i}$ of Eq.~(\ref{eq:phibreak}) becomes a constant field during $\Delta t$.
\HL{To summarize, the time-dependant variables on the right-hand side of Eq.~(\ref{eq:batcheler1967}) become $T_{i}$, and $T_{j}$;} 
\HL{hereinafter, we denote the temperature $T_{i}$ as $T_{i}(t)$ when we need to expressly show the time-dependency.}

\subsection{General case of $N$ particles}
We derive the form of entropy $S$ by integrating both sides of Eq.~(4) \HL{over time}:
\HL{
\begin{eqnarray}
\begin{split}
\int_{t_{0}}^{\tau}\frac{dS}{dt}dt &= \int_{t_{0}}^{\tau} \sum_{i}\frac{\phi_{i}}{T_{i}(t)} {dt} \\
		&+ \int_{t_{0}}^{\tau} \kappa \sum_{ij} \frac{F_{ij}}{d_{i}d_{j}T_{i}(t)T_{j}(t)} {T_{ij}(t)}^2 {dt}.
\end{split}
\label{eq:bothsideinteg1}
\end{eqnarray}
By changing the order of integration and summation, Eq.~(\ref{eq:bothsideinteg1}) can be rewritten as
\begin{eqnarray}
\begin{split}
\int_{t_{0}}^{\tau}\frac{dS}{dt}dt &=  \sum_{i} \phi_{i}\int_{t_{0}}^{\tau} \frac{1}{T_{i}(t)} {dt} \\
		&+ \kappa \sum_{ij} \frac{F_{ij}}{d_{i}d_{j}} \int_{t_{0}}^{\tau} \frac{T_{ij}(t)^2}{T_{i}(t)T_{j}(t)} {dt}.
\end{split}
\label{eq:bothsideinteg2}
\end{eqnarray}
The first term of Eq.~(\ref{eq:bothsideinteg2}) can be rewritten using the definition of definite integration by substitution~\cite{apostol1964mathematical} as
\begin{eqnarray}
\begin{split}
\int_{t_{0}}^{\tau}\frac{dS}{dt}dt &=  \sum_{i} \phi_{i}\int_{T_{i}^{0}}^{T_{i}^{\tau}} \frac{1}{T_{i}}\frac{dT_{i}}{dt} {dt} \nonumber \\
		&+ \kappa \sum_{ij} \frac{F_{ij}}{d_{i}d_{j}} \int_{t_{0}}^{\tau} \frac{T_{ij}(t)^2}{T_{i}(t)T_{j}(t)} {dt}, \nonumber \\
\end{split}
\end{eqnarray}
\begin{eqnarray}
\begin{split}
\raisebox{.2ex}{.}\raisebox{1.2ex}{.}\raisebox{.2ex}{.}
\int_{t_{0}}^{\tau}\frac{dS}{dt}dt &=  \sum_{i} \phi_{i}\int_{T_{i}^{0}}^{T_{i}^{\tau}} \frac{1}{T_{i}}dT_{i} \\
		&+ \kappa \sum_{ij} \frac{F_{ij}}{d_{i}d_{j}} \int_{t_{0}}^{\tau} \frac{T_{ij}(t)^2}{T_{i}(t)T_{j}(t)} {dt}.
\end{split}
\label{eq:bothsideinteg3}
\end{eqnarray}
}
\paragraph{Note added in (2026/02/25): Clarification on the substitution leading to Eq.~(\ref{eq:bothsideinteg3})}
{\it A minor notational ambiguity appears in the intermediate step following Eq.~(\ref{eq:bothsideinteg2}) concerning the substitution of the time integral by an integral over temperature.
Assuming that $T_i(t)$ is differentiable and monotonic on $[t_0,\tau]$, a mathematically consistent form of the substitution reads
\begin{equation}
\int_{t_0}^{\tau} \frac{1}{T_i(t)} \, dt
=
\int_{T_i^0}^{T_i^\tau}
\frac{1}{T_i}
\left( \frac{dT_i}{dt} \right)^{-1}
\, dT_i . \nonumber 
\end{equation}

In the infinitesimal time interval considered in the derivation, one may consistently adopt the same approximation used later in the manuscript and assume
\begin{equation}
\frac{dT_i}{dt} \simeq K = \mathrm{const.} \nonumber
\end{equation}
with $K \neq 0$. Accordingly, it is convenient to absorb the constant factor $1/K$ into an effective dissipation term
\begin{equation}
\bar{\phi}_i := \frac{\phi_i}{K}, \nonumber
\end{equation}
where $\bar{\phi}_i$ has units of entropy ($\mathrm{J\,K^{-1}}$).
Hereafter, for notational simplicity, we drop the bar and write $\phi_i$ instead of $\bar{\phi}_i$ in the integrated expressions.
The same notational point appears in the published journal version of the paper; the clarification given here applies equally to that version.
Importantly, this does not affect the structure of the subsequent derivations or the main conclusions of the work.}

\HL{
To integrate the second term in Eq.~(\ref{eq:bothsideinteg3}), we introduce the concept of general topology. Let us consider the following relationship between a differentiable function $\vec{G}(\vec{r})$ from some open subset $U$ ($\mathbb{R}^{n}$) to $\mathbb{R}$ and a differentiable function $\vec{r}$ from some closed interval to $U$. Then, by the multi-variable chain rules~\cite{williamson2multivariable}, we get
\begin{eqnarray}
\vec{G}(\vec{r})\cdot d\vec{r} &=& \vec{G}(\vec{r}(t))\cdot \frac{d \vec{r}(t)}{dt} dt.
\label{eq:substdiff}
\end{eqnarray} 
}
\HL{
Subsequently, we can integrate both sides of Eq.~(\ref{eq:substdiff}) to get
\begin{eqnarray}
\int \vec{G}(\vec{r})\cdot d\vec{r} &=& \int \vec{G}(\vec{r}(t))\cdot \frac{d \vec{r}(t)}{dt} dt.
\label{eq:substinteg}
\end{eqnarray} 
}
\HL{
Let us discuss the physical interpretations of the small element $d\vec{r}$ and the domain of integration when we set $\vec{r}$ to be $(T_{i},T_{j})$.
Because we perform the integration over a small time interval, we can regard $dT_{i}/dt$ and $dT_{j}/dt$ as constant values in the first approximation level. Due to the symmetricity of $i$ and $j$ in time dependence, the relationship of $dT_{i}/dt = dT_{j}/dt = K = {\rm const}$ can be established. Hence, Eq.~(\ref{eq:substinteg}) can be expressed using the vector $\vec{u} = (1,1)$ as follows:
\begin{eqnarray}
 \int \vec{G}(\vec{r}(t))\cdot\vec{u}~dt &=& \frac{1}{K}\int \vec{G}(\vec{r})\cdot d\vec{r}.
\end{eqnarray}
}
\HL{
Meanwhile, we can choose $\vec{G}(\vec{r})$ as
\begin{eqnarray}
	\vec{G}(\vec{r}(t))~~=~~G(\vec{r})\vec{u}~~=~~\frac{T_{ij}(t)^\HLL{2}}{T_{i}(t)T_{j}(t)}\vec{u} 
\end{eqnarray}
Although $\vec{G}$ includes a singular point at $(T_{i}, T_{j}) = (0, 0)$, we can exclude this point from the domain of integration because temperature in Kelvin is always positive.
We then obtain the following relationship:
\begin{eqnarray}
\int \frac{T_{ij}(t)^2}{T_{i}(t)T_{j}(t)} {dt} &=& \frac{1}{2K} \int \vec{G}(\vec{r})\cdot d\vec{r}. 
\label{eq:areaintegzero}
\end{eqnarray}
Here, we use the relationship of $\vec{u}\cdot\vec{u}={|u|}^{2}=2$.
}
\HL{
In the case that we integrate the left-hand side of Eq.~(\ref{eq:areaintegzero}) with respect to time $t$ in the range $[t_{0}, \tau]$, the corresponding domains of integration on the right-hand side with respect to $T_{i}$ and $T_{j}$ are $[T_{i}^{0}, T_{i}^{\tau}]$ and $[T_{j}^{0}, T_{j}^{\tau}]$, respectively. Besides, projection of the function $\vec{G}$ onto the $T_{i}T_{j}$-plane forms a finite area. Therefore, we regard the right-hand side as a surface integration with surface element vector $d\vec{r}$ as
}
\HL{
\begin{eqnarray}
\int_{t_{0}}^{\tau} \frac{T_{ij}(t)^2}{T_{i}(t)T_{j}(t)} {dt} &=& \frac{1}{2K} \int_{S} \vec{G}(\vec{r})\cdot d\vec{r}.
\label{eq:areaintegapprox}
\end{eqnarray}
Here, the notation of $S$ means that we perform the integration on the surface $S$ on the $T_{i}T_{j}$-plane in the ranges $[T_{i}^{0}, T_{i}^{\tau}]$ and $[T_{j}^{0}, T_{j}^{\tau}]$.
}

\HL{
Recall that we assume forward-Euler time integration in explicit simulations. 
Because all the known information we can use is defined at time $t_{0}$, we use the surface element vector $d\vec{r}^{0}$ as an alternative to $d\vec{r}$, as follows:
\begin{eqnarray}
\int_{t_{0}}^{\tau} \frac{T_{ij}(t)^2}{T_{i}(t)T_{j}(t)} {dt} &\simeq& \frac{1}{2K} \int_{S} \vec{G}(\vec{r})\cdot d\vec{r}^{0}.
\label{eq:areaintegapprox}
\end{eqnarray}
We then convert the right-hand side of Eq.~(\ref{eq:areaintegapprox}) into a double integration on the $T_{i}T_{j}$-plane according to the formula for a surface integral of a scalar function $G$ over a surface~$S$~\cite{hazewinkel2001gram, leathem1922volume}, as
\begin{eqnarray}
&~& \frac{1}{2K} \int_{S} \vec{G}(\vec{r})\cdot d \vec{r}^{0} \\
&=& \frac{1}{2K} \int_{S} G(\vec{r})\vec{u}\cdot\vec{u}d{\rm r}^{0} \\
&=& \frac{1}{K} \int_{S} G(\vec{r})d{\rm r}^{0} \\
&=& \frac{1}{K} \int_{T_{j}^{0}}^{T_{j}^{\tau}} \!\!\! \int_{T_{i}^{0}}^{T_{i}^{\tau}} G(T_{i},T_{j}) \Biggr|\frac{\partial \vec{r^{0}}}{\partial T_{i}}\times\frac{\partial \vec{r^{0}}}{\partial T_{j}}\Biggl|~dT_{i} dT_{j},
\label{eq:areaintegsurface}
\end{eqnarray}
}
\HL{
where 
\begin{eqnarray}
\Biggr|\frac{\partial \vec{r^{0}}}{\partial T_{i}}\times\frac{\partial \vec{r^{0}}}{\partial T_{j}}\Biggl| 
	&=& \Biggr(\frac{\partial T_{i}^{0}}{\partial T_{i}}\Biggl) \Biggr(\frac{\partial T_{j}^{0}}{\partial T_{j}}\Biggl)
	- \Biggr(\frac{\partial T_{i}^{0}}{\partial T_{j}}\Biggl) \Biggr(\frac{\partial T_{j}^{0}}{\partial T_{i}}\Biggl).
\label{eq:TiTjand2dcross}
\end{eqnarray}
}
\HL{
Each element of the right-hand side in Eq.~(\ref{eq:TiTjand2dcross}) represents the gradient of $T_{i}$ or $T_{j}$ on the $T_{i}T_{j}$-plane at time $t_{0}$.
Similarly to the aforementioned case of their time differential, we approximate these gradients as constant. Thus, Eq.~(\ref{eq:areaintegapprox}) can be rewrittened using 
Eq.~(\ref{eq:areaintegsurface}), Eq.~(\ref{eq:TiTjand2dcross}), and a constant value $M$ as follows:
\begin{eqnarray}
\int_{t_{0}}^{\tau} \frac{T_{ij}(t)^2}{T_{i}(t)T_{j}(t)} {dt} 
&=& \frac{1}{K} \int_{T_{j}^{0}}^{T_{j}^{\tau}} \!\!\! \int_{T_{i}^{0}}^{T_{i}^{\tau}} G(T_{i},T_{j}) M~dT_{i} dT_{j}, \nonumber \\
&=& \frac{M}{K} \int_{T_{j}^{0}}^{T_{j}^{\tau}} \!\!\! \int_{T_{i}^{0}}^{T_{i}^{\tau}} G(T_{i},T_{j}) dT_{i} dT_{j}, \label{eq:TiTjandfin}
\end{eqnarray}
where
\begin{eqnarray}
M = \Biggr|\frac{\partial \vec{r^{0}}}{\partial T_{i}}\times\frac{\partial \vec{r^{0}}}{\partial T_{j}}\Biggl| =~{\rm const}.
\end{eqnarray}
}
\HL{By substituting Eq.~(\ref{eq:Tijbreak}) and Eq.~(\ref{eq:TiTjandfin}) into Eq.~(\ref{eq:bothsideinteg3}), we obtain }
\begin{eqnarray}
\begin{split}
\int_{t_{0}}^{\tau}\frac{dS}{dt}dt &=  \sum_{i} \phi_{i}\int_{T_{i}^{0}}^{T_{i}^{\tau}} \frac{1}{T_{i}}{dT}_{i}\\
		&+ \bar{\kappa} \sum_{ij} \frac{F_{ij}}{d_{i}d_{j}} \int_{T_{j}^{0}}^{T_{j}^{\tau}} \!\!\! \int_{T_{i}^{0}}^{T_{i}^{\tau}} \frac{{(T_{i}-T_{j})}^2}{T_{i}T_{j}} {dT}_{i}{dT}_{j}. 
\end{split}
\label{eq:batcheler1967:integral:second}
\end{eqnarray}
\HL{Here, $\bar{\kappa} = \kappa M/K$. For simplicity, in this paper, we confine $\bar{\kappa}$ to the case of $\bar{\kappa} > 0 $. }

By integrating the left-hand side of Eq.~(\ref{eq:batcheler1967:integral:second}), we obtain
\begin{eqnarray}
\begin{split}
S(\tau) &=  S(t_{0}) + \sum_{i} \phi_{i}\int_{T_{i}^{0}}^{T_{i}^{\tau}} \frac{1}{T_{i}}{dT}_{i}\\
		&+ \bar{\kappa} \sum_{ij} \frac{F_{ij}}{d_{i}d_{j}} \int_{T_{j}^{0}}^{T_{j}^{\tau}} \!\!\! \int_{T_{i}^{0}}^{T_{i}^{\tau}} \frac{{(T_{i}-T_{j})}^2}{T_{i}T_{j}} {dT}_{i}{dT}_{j}. \\
\end{split}
\label{eq:batcheler1967:integral:fifth}
\end{eqnarray}
Here, $S(t_{0})$ indicates the initial entropy at time $t_{0}$.
Finally, by performing \HL{integration of the second term with respect to $T_{i}$ and} double integration of the third term on the right-hand side of Eq.~(\ref{eq:batcheler1967:integral:fifth}) with respect to 
$T_{i}$ and $T_{j}$, we obtain the form of entropy of SDPD at time $\tau$ for the general case of $N$ particles:
\begin{eqnarray}
\begin{split}
S(\tau) &=  S(t_{0})~+~\sum_{i} \phi_{i}{\rm ln}\Biggl(\frac{T_{i}^{\tau}}{T_{i}^{0}}\Biggl)~+~\frac{\bar{\kappa}}{2} \sum_{ij} \frac{F_{ij}}{d_{i}d_{j}} O_{ij},\\
O_{ij}  &= \left\{{\left(T_{i}^{\tau}\right)}^2-{\left(T_{i}^{0}\right)}^2 \right\}{\rm ln} \Biggl(\frac{T_{j}^{\tau}}{T_{j}^{0}}\Biggl)\\
		&+ \left\{{\left(T_{j}^{\tau}\right)}^2-{\left(T_{j}^{0}\right)}^2 \right\}{\rm ln} \Biggl(\frac{T_{i}^{\tau}}{T_{i}^{0}}\Biggl)\\
	   	&- 4\left({T_{i}^{\tau}}-{T_{i}^{0}}\right) \left({T_{j}^{\tau}}-{T_{j}^{0}}\right).
\end{split}
\label{eq:batcheler1967:integral:sixth}
\end{eqnarray}

\subsection{Specific case of two particles}
When $N=2$, Eq.~(\ref{eq:batcheler1967:integral:sixth}) can be written as
\begin{eqnarray}
\begin{split}
S(\tau) &=  S(t_{0})+\phi_{1}{\rm ln}\Biggl(\frac{T_{1}^{\tau}}{T_{1}^{0}}\Biggr) + \phi_{2}{\rm ln}\Biggl(\frac{T_{2}^{\tau}}{T_{i2}^{0}}\Biggr)\\
		&+ \Biggl(\frac{\bar{\kappa} F_{12}}{2 d_{1}d_{2}} + \frac{\bar{\kappa} F_{21}}{2d_{2}d_{1}} \Biggl) 
		\left\{{\left(T_{1}^{\tau}\right)}^2-{\left(T_{1}^{0}\right)}^2 \right\}{\rm ln} \Biggl(\frac{T_{2}^{\tau}}{T_{2}^{0}}\Biggl)\\
		&+ \Biggl(\frac{\bar{\kappa} F_{12}}{2 d_{1}d_{2}} + \frac{\bar{\kappa} F_{21}}{2d_{2}d_{1}} \Biggl) 
		\left\{{\left(T_{2}^{\tau}\right)}^2-{\left(T_{2}^{0}\right)}^2 \right\}{\rm ln} \Biggl(\frac{T_{1}^{\tau}}{T_{1}^{0}}\Biggl)\\
	   	&- 4\Biggl(\frac{\bar{\kappa} F_{12}}{2 d_{1}d_{2}} + \frac{\bar{\kappa} F_{21}}{2d_{2}d_{1}} \Biggl)\left({T_{1}^{\tau}}-{T_{1}^{0}}\right) \left({T_{2}^{\tau}}-{T_{2}^{0}}\right). 
		\label{eq:entropy:n2:first}
\end{split}
\end{eqnarray}
Because the symmetries of $d_{i}$ and $F_{ij}$~$(i=1,2)$ can be confirmed from Eq.~(\ref{eq:dijbreak}) and Eq.~(\ref{eq:Fijbreak}) in the case of $N=2$, 
the relationships $d_{\HL{1}}=d_{\HL{2}}$ and $F_{12}=F_{21}$ are true.
Hence, the constant parameters of $d$, $F$, and $\alpha$ can be introduced as
\begin{eqnarray}
\alpha&:=&\frac{\bar{\kappa} F}{d^{2}}\label{eq:alpha},\\
     d&:=&d_{\HL{1}}=d_{\HL{2}} \label{eq:d:def},\\
     F&:=&F_{12}=F_{21} \label{eq;F:def}.
\end{eqnarray}
Likewise, the viscous heating fields of $\phi_{1}$ and $\phi_{2}$ become equal. 
The parameter $\phi$ can be introduced as
\begin{eqnarray}
\begin{split}
\phi &= \phi_{1} = \phi_{2}, \\
	\phi &= \frac{\alpha}{\bar{\kappa}}\Biggl( \frac{5\eta}{6}-\frac{\zeta}{2}\Biggr){\left(\vec{v}_{ij}\right)}^{2} 
	     + \frac{5\alpha}{2\bar{\kappa}}\Biggl( \zeta + \frac{\eta}{3}\Biggr){\left(\vec{e}_{ij} \cdot \vec{v}_{ij}\right)}^{2}. 
\end{split}
\label{eq;phi:def}
\end{eqnarray}
Let us recall that the vectors of $\vec{e}_{ij}$ and $\vec{v}_{ij}$ become constant between $t_{0}$ and $t_{0} + \Delta t$.
By using the parameters from Eq.~(\ref{eq:alpha}) to Eq.~(\ref{eq;phi:def}), Eq.~(\ref{eq:entropy:n2:first}) can be simply rewritten as
\begin{eqnarray}
\begin{split}
S(\tau) &=  S(t_{0})\\ 
		&+ \Biggl[\phi+\alpha\left\{{\left(T_{1}^{\tau}\right)}^2-{\left(T_{1}^{0}\right)}^2 \right\}\Biggr]{\rm ln} \Biggl(\frac{T_{2}^{\tau}}{T_{2}^{0}}\Biggl)\\
		&+ \Biggl[\phi+\alpha\left\{{\left(T_{2}^{\tau}\right)}^2-{\left(T_{2}^{0}\right)}^2 \right\}\Biggr]{\rm ln} \Biggl(\frac{T_{1}^{\tau}}{T_{1}^{0}}\Biggl)\\
	   	&- 4\alpha\left({T_{1}^{\tau}}-{T_{1}^{0}}\right) \left({T_{2}^{\tau}}-{T_{2}^{0}}\right). \\
		\label{eq:entropy:n2:second}
\end{split}
\end{eqnarray}
\subsection{Characteristics of the parameter $\alpha$}
\begin{figure}[t]
\centerline{\includegraphics[scale=0.3]{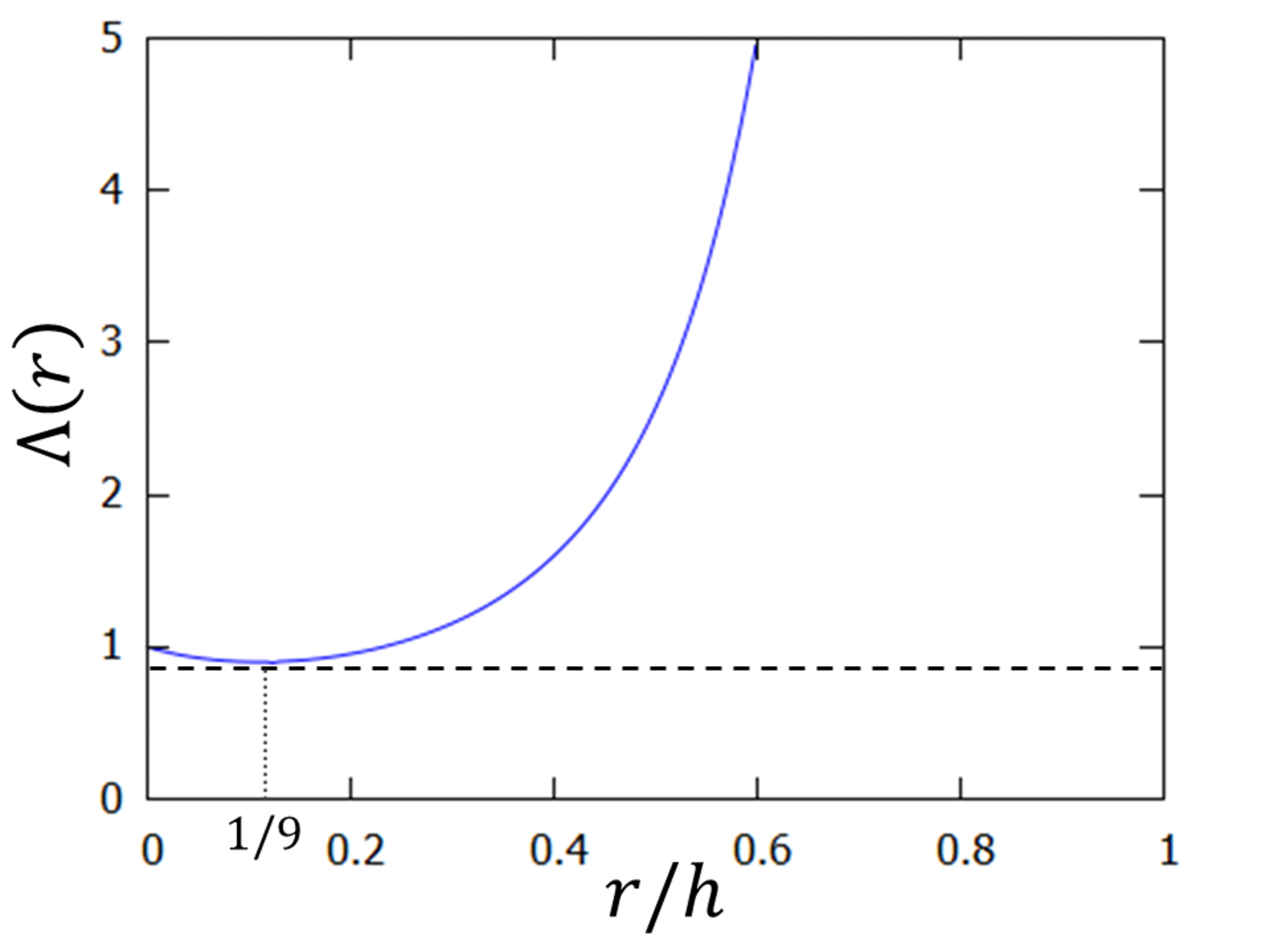}}
\caption{Function of $\Lambda(r)$.}
\label{fig:func:lambdar}
\end{figure}

As a foundation for the next section, let us examine the characteristics of the parameter $\alpha$.
The detailed expression of Eq.~(\ref{eq:alpha}) is written as
\begin{eqnarray}
\begin{split}
\alpha &= C_{\alpha} \Lambda(r), \\
C_{\alpha} &= \bar{\kappa}{\Biggl(\frac{315}{4\pi h^5}\Biggr)}{\Biggl(\frac{105}{16 \pi h^3}\Biggr)}^{-2} = {\rm const}., \\
\Lambda(r) &= \frac{\Bigl(1-\frac{r}{h}\Bigr)^2}{{\Bigl(1+3\frac{r}{h}\Bigr)}^2 \Bigl(1-\frac{r}{h}\Bigr)^6}.
\end{split}
\label{eq:alpha:breakdown}
\end{eqnarray}
Here, $C_{\alpha}> 0$ because $\bar{\kappa} > 0$. Figure~\ref{fig:func:lambdar} shows the function of $\Lambda(r)$ in Eq.~(\ref{eq:alpha:breakdown}). Here, $\Lambda(r)$ is confirmed to become an increasing function after $r/h > 1/9$, and it \HLL{is} larger than zero within the range of $0\le~r/h~\le1$.
Hence, the resulting $\alpha$ becomes positive within $0\le~r/h~\le1$.
\HL{Although} the function $\Lambda(x)$ becomes discontinuous for $r/h \gg 1$, the region of $r/h \gg 1$ is not referred to in the SDPD, so this is not a problem.

In the next section, we discuss the characteristics of a two-particle system of SDPD regarding the thermodynamic validity by examining entropy stability conditions.

\section{Entropy stability analysis}\label{seq:esa}
\subsection{Derivation of entropy stability conditions}
Denote the number of particles as $N$, the volume of the system as $V$, and the internal energy as $U$.
Under the conditions of a constant volume process and fixed number of particles 
($N=V={\rm const.}$), \HL{an entropy stability condition is derived from a thought experiment in which two identical systems transfer an infinitesimal internal energy $\Delta U$ between each other. If the entropy is stable, it does not increase regardless of the transfer~(namely, $S(U-\Delta U) + S(U+\Delta U) < 2S(U)$). By taking the Taylor series expansion of both sides of this relationship, we obtain the following stability condition~\cite{Callen:450289}:}
\begin{eqnarray}
\frac{\partial^2 S}{\partial U\HL{^2}}\HL{\Biggr|_{V,N}} \le 0. \label{eq:stability-cond1}
\end{eqnarray}

Let us impose the condition of Eq.~(\ref{eq:stability-cond1}) on the entropy $S$ in Eq.~(\ref{eq:entropy:n2:second}). 
The left-hand side of Eq.~(\ref{eq:stability-cond1}), i.e., the second-order partial derivatives of the entropy $S$ by the internal energy $U$, is calculated by the following processes. First, we derive the first-order partial derivatives of $S$ from the temperatures $T_{1}^{\tau}$ and $T_{2}^{\tau}$ as 
\begin{eqnarray}
\begin{split}
\frac{\partial S}{\partial T_{1}^{\tau}} 
&= 2\alpha T_{1}^{\tau}{\rm ln} \Biggl(\frac{T_{2}^{\tau}}{T_{2}^{0}}\Biggl) \\
&+ \frac{\Biggl[\phi+\alpha\left\{{\left(T_{2}^{\tau}\right)}^2-{\left(T_{2}^{0}\right)}^2 \right\}\Biggr]}{T^{\HL{\tau}}_{1}} \\
&- 4\alpha\left({T_{2}^{\tau}}-{T_{2}^{0}}\right), 
\end{split}
\label{eq:stability:second:first}
\end{eqnarray}
\begin{eqnarray}
\begin{split}
\frac{\partial S}{\partial T_{2}^{\tau}} 
&= 2\alpha T_{2}^{\tau}{\rm ln} \Biggl(\frac{T_{1}^{\tau}}{T_{1}^{0}}\Biggl) \\
&+ \frac{\Biggl[\phi+\alpha\left\{{\left(T_{1}^{\tau}\right)}^2-{\left(T_{1}^{0}\right)}^2 \right\}\Biggr]}{T^{\HL{\tau}}_{2}} \\
&- 4\alpha\left({T_{1}^{\tau}}-{T_{1}^{0}}\right), 
\end{split}
\end{eqnarray}
\begin{eqnarray}
\begin{split}
 \frac{\partial }{\partial T_{2}^{\tau}} \Biggl(\frac{\partial S}{\partial T_{1}^{\tau}}\Biggr) 
&= 2\alpha\Biggl( \frac{T_{1}^{\tau}}{T_{2}^{\tau}} + \frac{T_{2}^{\tau}}{T_{1}^{\tau}} - 2 \Biggr),
\end{split}
\end{eqnarray}
\begin{eqnarray}
\begin{split}
 \frac{\partial }{\partial T_{1}^{\tau}} \Biggl(\frac{\partial S}{\partial T_{2}^{\tau}}\Biggr) 
&= 2\alpha\Biggl( \frac{T_{2}^{\tau}}{T_{1}^{\tau}} + \frac{T_{1}^{\tau}}{T_{2}^{\tau}} - 2 \Biggr),
\end{split}
\end{eqnarray}
\begin{eqnarray}
\frac{\partial }{\partial T_{2}^{\tau}} \Biggl(\frac{\partial S}{\partial T_{1}^{\tau}}\Biggr)
&=&	
\frac{\partial }{\partial T_{1}^{\tau}} \Biggl(\frac{\partial S}{\partial T_{2}^{\tau}}\Biggr), 
\end{eqnarray}
\begin{eqnarray}
\begin{split}
\frac{\partial }{\partial T_{1}^{\tau}} \Biggl(\frac{\partial S}{\partial T_{1}^{\tau}}\Biggr)
&= 2\alpha {\rm ln} \Biggl(\frac{T_{2}^{\tau}}{T_{2}^{0}}\Biggl) \\
&+ \frac{-\Biggl[\phi+\alpha\left\{{\left(T_{2}^{\tau}\right)}^2-{\left(T_{2}^{0}\right)}^2 \right\}\Biggr]}{{T_{1}^{\HL{\tau}}}^2}, 
\end{split}
\end{eqnarray}
\begin{eqnarray}
\begin{split}
\frac{\partial }{\partial T_{2}^{\tau}} \Biggl(\frac{\partial S}{\partial T_{2}^{\tau}}\Biggr)
&= 2\alpha {\rm ln} \Biggl(\frac{T_{1}^{\tau}}{T_{1}^{0}}\Biggl) \\
&+ \frac{-\Biggl[\phi+\alpha\left\{{\left(T_{1}^{\tau}\right)}^2-{\left(T_{1}^{0}\right)}^2 \right\}\Biggr]}{{T_{2}^{\HL{\tau}}}^2}. 
\end{split}
\end{eqnarray}
The heat capacity $C_{V}$ under an isochoric process is given by
\begin{eqnarray}
\frac{\partial T_{i}^{\tau}}{\partial U}\HL{\Biggr|_{V}} = \frac{dT_{i}^{\tau}}{dU}\HL{\Biggr|_{dV=0}} = \frac{1}{C_{V}}={\rm const.}~~(i=1,2). \label{eq:heatcap}
\label{eq:stability:second:last}
\end{eqnarray}
Meanwhile, the left-hand side of Eq.~(\ref{eq:stability-cond1}) can be rewritten as follows~\cite{Callen:450289}:
\begin{eqnarray}
\begin{split}
\frac{\partial^2 S}{\partial U^\HL{2}}\HL{\Biggr|_{V,N} }
&= \Biggl(\frac{\partial^2 S}{\partial {T_{1}^{\tau}}^2}\Biggr)\Biggl(\frac{\HL{d} T_{1}^{\tau}}{\HL{d} U}\Biggr)^2 \\
&+ 2\frac{\partial }{\partial T_{1}^{\tau}} \Biggl(\frac{\partial S}{\partial T_{2}^{\tau}}\Biggr)\Biggl(\frac{dT_{1}^{\tau}}{dU}\Biggr) \Biggl(\frac{dT_{2}^{\tau}}{dU}\Biggr) \\
&+ \Biggl(\frac{\partial^2 S}{\partial {T_{2}^{\tau}}^2}\Biggr)\Biggl(\frac{\HL{d} T_{2}^{\tau}}{\HL{d} U}\Biggr)^2 \\
&+ \HL{\Biggl(\frac{\partial S}{\partial {T_{1}^{\tau}}}\Biggr)\Biggl(\frac{d^2 T_{1}^{\tau}}{d U^2}\Biggr)}
 + \HL{\Biggl(\frac{\partial S}{\partial {T_{2}^{\tau}}}\Biggr)\Biggl(\frac{d^2 T_{2}^{\tau}}{d U^2}\Biggr)}.
\end{split}
\label{eq:stability:detail1}
\end{eqnarray}
\HL{Here, the fourth and fifth terms vanish because of Eq.~(\ref{eq:stability:second:last}).} 
Using Eq.~(\ref{eq:stability:second:first}) to Eq.~(\ref{eq:stability:detail1}), we obtain the following entropy stability condition: 
\begin{eqnarray}
\begin{split}
&\frac{1}{{C_{V}}^2}\Biggl[ 
   \frac{-\Biggl\{\phi+\alpha\left({\left(T_{2}^{\tau}\right)}^2-{\left(T_{2}^{0}\right)}^2 \right)\Biggr\}}{{T_{1}^{\HL{\tau}}}^2} \\ 
&+ \frac{-\Biggl\{\phi+\alpha\left({\left(T_{1}^{\tau}\right)}^2-{\left(T_{1}^{0}\right)}^2 \right)\Biggr\}}{{T_{2}^{\HL{\tau}}}^2} \\ 
&+ 4\alpha\Biggl\{     
		  \frac{1}{2}{\rm ln} \Biggl(\frac{T_{1}^{\tau}}{T_{1}^{0}}\Biggl) 
		+ \frac{1}{2}{\rm ln} \Biggl(\frac{T_{2}^{\tau}}{T_{2}^{0}}\Biggl) 
		+ \frac{T_{2}^{\tau}}{T_{1}^{\tau}} + \frac{T_{1}^{\tau}}{T_{2}^{\tau}} -2 \Biggr\}
	\Biggr] \le 0,
\end{split}
\label{eq:stability:detail2}
\end{eqnarray}
\begin{eqnarray}
\nonumber
\end{eqnarray}
\begin{eqnarray}
\begin{split}
&\raisebox{.2ex}{.}\raisebox{1.2ex}{.}\raisebox{.2ex}{.}
~~~\frac{\phi+\alpha\left({\left(T_{2}^{\tau}\right)}^2-{\left(T_{2}^{0}\right)}^2 \right)}{{T_{1}^{\HL{\tau}}}^2} \\ 
&+ \frac{\phi+\alpha\left({\left(T_{1}^{\tau}\right)}^2-{\left(T_{1}^{0}\right)}^2 \right)}{{T_{2}^{\HL{\tau}}}^2} \\ 
&-4\alpha\Biggl\{ 
		  \frac{1}{2}{\rm ln} \Biggl(\frac{T_{1}^{\tau}}{T_{1}^{0}}\Biggl) 
	    + \frac{1}{2}{\rm ln} \Biggl(\frac{T_{2}^{\tau}}{T_{2}^{0}}\Biggl) 
		+ \frac{T_{2}^{\tau}}{T_{1}^{\tau}} + \frac{T_{1}^{\tau}}{T_{2}^{\tau}} - 2 \Biggr\} \ge 0.
\end{split}
\label{eq:stability:detail3}
\end{eqnarray}
\\

\begin{figure}[t]
\centerline{\includegraphics[scale=0.33]{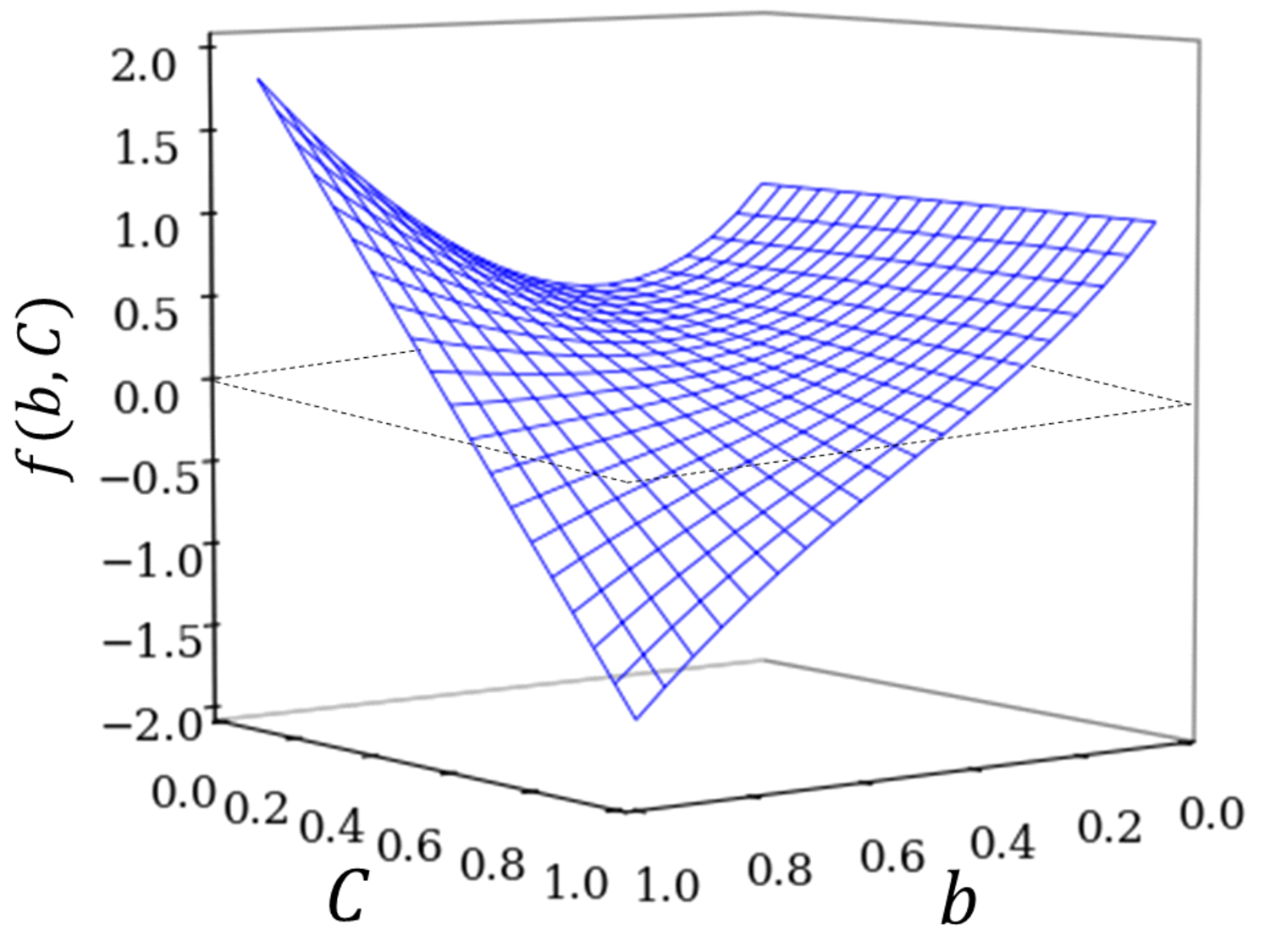}}
\caption{Fourth-order function of $f(b,C)$ described by Eq.~(\ref{eq:4thorderfunc}).}
\label{fig:4thorderfunc}
\end{figure}
\subsection{Theoretical analyses with case studies}
\HL{To reproduce a system with a constant temperature gradient, we impose a constraint on the domain of integration as follows:}
\begin{eqnarray}
\begin{split}
T_{2}^{\tau} = b T_{1}^{\tau},~~T_{2}^{0} = b T_{1}^{0},~~b > 0.\\
\end{split}
\label{eq:stablecond:linear}
\end{eqnarray}
\HL{Equation~(\ref{eq:stablecond:linear}) suggests that a linear temperature relationship is imposed before and after the time evolution.}
By substituting the right-hand side of the two equations in Eq.~(\ref{eq:stablecond:linear}) into Eq.~(\ref{eq:stability:detail3}),
we obtain 
\begin{eqnarray}
\alpha f(b,C)T_{1}^{\tau} \ge \alpha(b^4+1)T_{1}^{0}-\phi(b^2+1),
\label{eq:stablecondvnezero}
\end{eqnarray}
where 
\begin{eqnarray}
\begin{split}
f(b,C) &= b^4-4b^3\\
	   &+(8-4C)b^2-4b+1,\\
C &:= {\rm ln}\Biggl(\frac{T_{1}^{\tau}}{T_{1}^{0}}\Biggr).
\end{split}
\label{eq:4thorderfunc}
\end{eqnarray}
Because $\alpha > 0$, the sign of the left-hand side of Eq.(\ref{eq:stablecondvnezero}) is determined by the sign of the function $f(b,C)$.
Figure~\ref{fig:4thorderfunc} shows a contour plot of $f(b,C)$. 
The function $f(b,C)$ increases as the parameter $b$ increases when $C=0$, whereas it decreases as $b$ increases when $C \gg 0$. 
It is confirmed 
that there exist critical values $C_{c}$ where $f(b,C)=0$.
Therefore, the stability condition is distinguished by the function $f(b,C)$ as
\begin{eqnarray}
\begin{cases}
f(b,C) \ge 0 & T_{1}^{\tau} \ge \frac{\alpha(b^4+1)T_{1}^{0}-\phi(b^2+1)}{\alpha f(b,C)}, \\
f(b,C) \le 0 & T_{1}^{\tau} \le \frac{\alpha(b^4+1)T_{1}^{0}-\phi(b^2+1)}{\alpha f(b,C)}.
\end{cases}
\label{eq:stability-cond-vnezero}
\end{eqnarray}
In the case that the parameter $b$ is sufficiently close to 1 (i.e., the temperature gradient between two particles is moderate), 
Eq.~(\ref{eq:stability-cond-vnezero}) works as a stabilizer of the system; 
it gives the maximum temperature limit as $T_{1}^{\tau}$ increases and reaches the high-temperature area where $f(b,C) \le 0$, whereas 
it gives the minimum temperature limit as $T_{1}^{\tau}$ decreases and reaches the low-temperature area where $f(b,C) \ge 0$.
Meanwhile, when the parameter $b$ is sufficiently close to 0 (i.e., the temperature gradient between two particles is steep), $f(b,C) $ becomes positive everywhere, and the upper part of Eq.~(\ref{eq:stability-cond-vnezero}) is therefore required as the stability condition.

\begin{figure}[t]
\centerline{\includegraphics[scale=0.3]{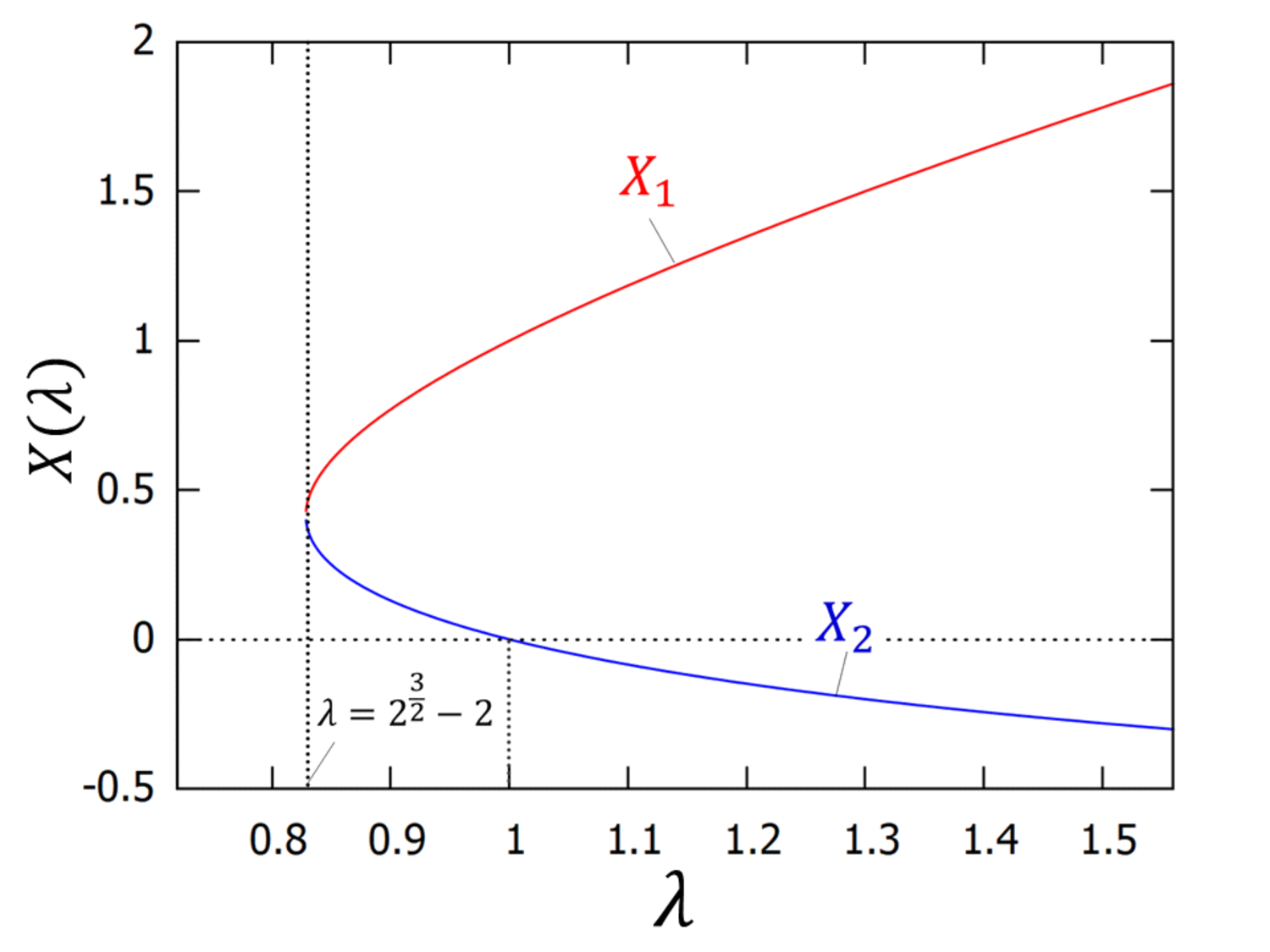}}
\caption{Dependencies of $X_1$ and $X_{2}$ on the parameter $\lambda$.}
\label{fig:x1x2dependslam}
\end{figure}
\begin{figure}[t]
\centerline{\includegraphics[scale=0.3]{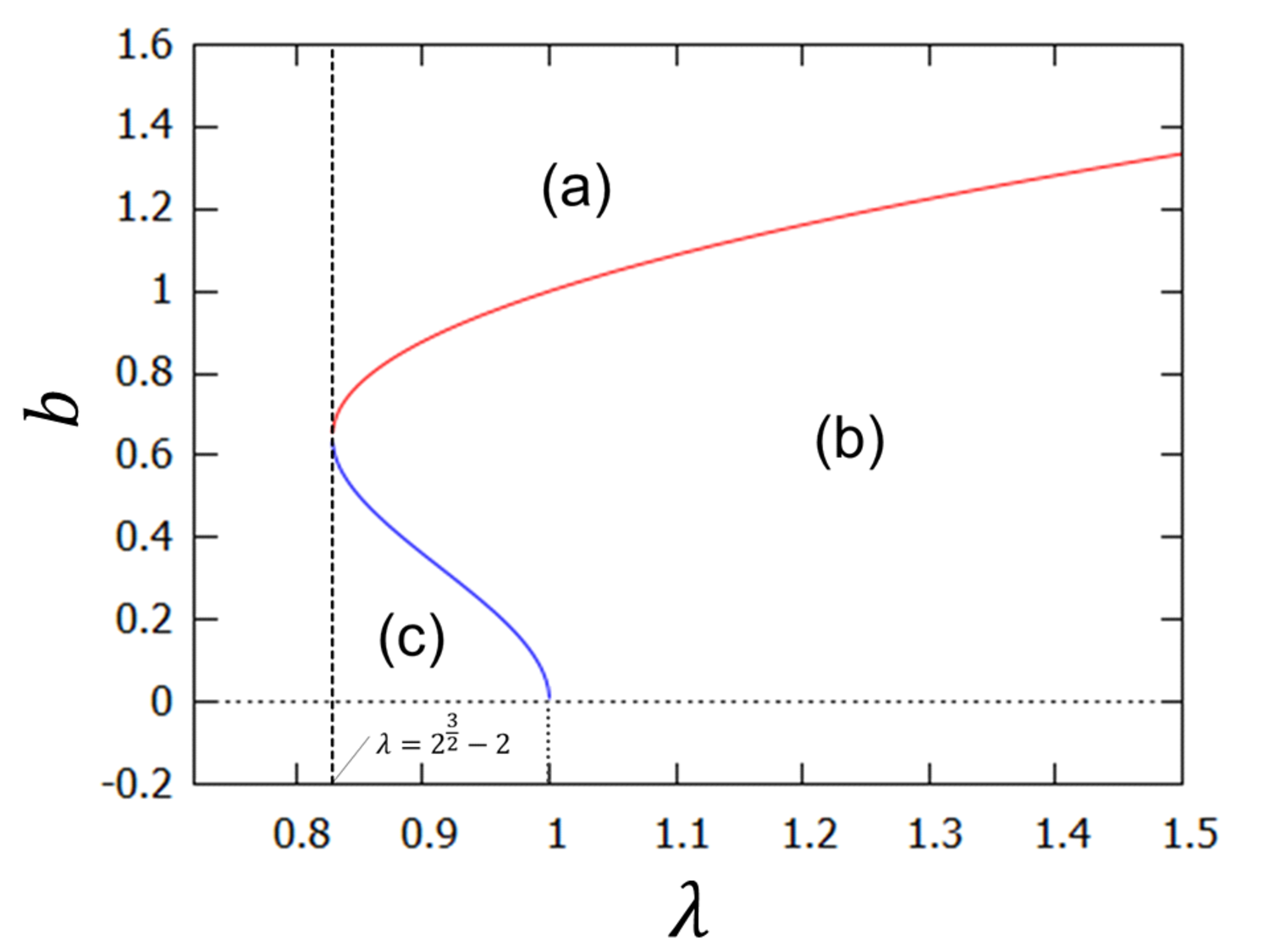}}
\caption{Classification of the state of the SDPD system determined by the parameters $b$ and $\lambda$ when $f(b,C)\ge0$.}
\label{fig:bdependslam}
\end{figure}
\renewcommand{\arraystretch}{1.8}
\begin{table*}[t]
\centering
\scalebox{2.0}[2.0]{
\tiny
\begin{tabular}{ccccc} 
\hline
Types& $f(b,C)$ & $F$ & $D$ & Entropy-stability Conditions  \\ 
\hline
1 & $+$ & $+$ & $+$ & $T_{1}^{\tau}\ge \frac{D}{\alpha f(b,C)}$ \\
2 & $+$ & $+$ & $-$ & $T_{1}^{\tau} > 0$ \\
3 & $+$ & $-$ & $+$ & (unstable) \\
4 & $+$ & $-$ & $-$ & $T_{1}^{\tau}\le \frac{D}{\alpha f(b,C)}$ \\
5 & $-$ & $+$ & $+$ & (unstable) \\
6 & $-$ & $+$ & $-$ & $T_{1}^{\tau}\le \frac{D}{\alpha f(b,C)}$ \\
7 & $-$ & $-$ & $+$ & $T_{1}^{\tau}\ge \frac{D}{\alpha f(b,C)}$ \\
8 & $-$ & $-$ & $-$ & $T_{1}^{\tau} > 0$ \\
\hline
\end{tabular}
}
\HL{\caption{Eight different types of entropy-stability conditions depending on function $f(b, C)$, the parameter $D$, and the function $F$.}}
\end{table*}
\renewcommand{\arraystretch}{1.0}

To deepen the discussion on this issue, we provide a visual representation of the case of $f(b,C) \ge 0$ in Eq.~(\ref{eq:stability-cond-vnezero}).
\HL{Denote the numerator of the fraction in Eq.~(\ref{eq:stability-cond-vnezero}) as $D$.} 
\HL{Let us examine the conditions that the parameter $D$ becomes positive as}
\begin{eqnarray}
D &=& \alpha(b^4+1)T_{1}^{0}-\phi(b^2+1) \ge 0. 
\end{eqnarray}
By replacing $b^2$ with the parameter $X$, we obtain
\begin{eqnarray}
\alpha(X^2+1)T_{1}^{0}-\phi(X+1) \ge 0,~~~X=b^2. \label{eq:cond:nume}
\end{eqnarray}
When the equality holds, we obtain the solutions of $X_1$ and $X_2$ as
\begin{eqnarray}
\begin{split}
X_1 &= \frac{\lambda+\sqrt{{{\lambda }^{2}}+4 \lambda -4} }{2}, \\
X_2 &= \frac{\lambda-\sqrt{{{\lambda }^{2}}+4 \lambda -4} }{2}, \\
\lambda &= \frac{\phi}{\alpha T_{1}^{0}}.
\end{split}
\label{eq:x1x2sol}
\end{eqnarray}
Figure~\ref{fig:x1x2dependslam} shows the dependencies of $X_{1}$ and $X_{2}$ on the parameter $\lambda$.
The solution of $X_{1}$ is valid when $\lambda \ge {{2}^{\frac{3}{2}}}-2$ and is confirmed to satisfy $X_{1} > X_{2}$ everywhere. 
Meanwhile, the solution of $X_{2}$ is valid only when ${{2}^{\frac{3}{2}}}-2 \le \lambda \le 1$. 
Hence, by substituting the expressions in Eq.~(\ref{eq:x1x2sol}) into Eq.~(\ref{eq:cond:nume}), we obtain the following conditions:
\begin{eqnarray}
\begin{split}
&b \ge \sqrt{\frac{\lambda+\sqrt{{{\lambda }^{2}}+4 \lambda-4} }{2}},~\lambda \ge \lambda_0, \\
&b \le \sqrt{\frac{\lambda-\sqrt{{{\lambda }^{2}}+4 \lambda-4} }{2}},~b > 0,~\lambda_0 \le \lambda \le 1, \\
&\lambda_0 := {{2}^{\frac{3}{2}}}-2.
\end{split}
\label{eq:bcond}
\end{eqnarray}

Figure~\ref{fig:bdependslam} shows the classification of the state of the SDPD system determined by the parameters $b$ and $\lambda$. 
\HL{In area (b), the sign of the parameter $D$ becomes negative. In contrast, in areas (a) and (c), the sign of $D$ becomes positive. From the results in Fig.~\ref{fig:x1x2dependslam} and Fig.~\ref{fig:bdependslam} it is important to note that the sign of the parameter $D$ could become \HLL{positive or negative}.
	
Given that the signs of $\alpha$ and the function $F$ are the same, we can subdivide the entropy stability conditions into eight different types, as listed in Table~1.}
\HL{In Type~2 and Type~8, the system becomes stable because the temperature in Kelvin must be positive, while the sign of $D/\alpha f(b, C)$ is negative. However, in Type~3 and Type~5, the entropy becomes unstable everywhere, and no scenario satisfies the condition that the temperature becomes less than or equal to $D/\alpha f(b, C)$ because it is negative.}

\HL{
The main point to emphasize from Table~1 is that the entropy stability condition of the system depends on the function $F$, which indicates that the types of kernel functions influence the entropy stability condition of the system. In the classical SDPD model, because it uses the Lucy kernel, the sign of the function $F$ becomes positive. Hence, the possible types are Type~1, Type~2, Type~5, and Type~6. 

Let us consider same cases using other kernel functions. In the case that we use the spiky kernel~\cite{desbrun1996smoothed} or poly6 kernel~\cite{muller2003particle}, each $F$ is positive everywhere in the range of $0 < r/h < 1$. Therefore, the possible types of entropy conditions are the same as those for the Lucy kernel. On the contrary, when we use Mao and Yang's spline kernel~\cite{mao2006particle, monaghan1992smoothed, becker2007weakly}, the possible types of entropy conditions are Type~3, Type~4, Type~7, and Type~8 because the function $F$ becomes negative in the range of $0<r/h<1$. For reference, Fig.~5 shows the function $F$ in different types of kernel functions. 
}

\HL{
The fact that the kernel function contributes to the entropy stability conditions of a two-particle system can be extended to many-particle systems based on the concept of pair-wise particle methods~\cite{yang2017pairwise, tartakovsky2005modeling, bandara2013smoothed, tartakovsky2016smoothed}. In these methods, the total force $\vec{f}$ acted on a particle is broken down as $\vec{f}=\sum \vec{f}_{ij}$, where $\vec{f}_{ij}$ is the force between the $i$th and $j$th particles~\cite{yang2017pairwise}. Namely, the dynamics of a total system can be described as superpositions of two-particle modes. 
}

\HL{
Consider a case that two persons carry out simulations using the same computational conditions except for their kernel functions. If one uses the spiky kernel and the other uses the spline kernel, the entropy stability conditions imposed on an identical pair of particles differ. In this case, it could happen that only one of the systems is judged to be unstable despite them considering the same physical scenario, which could lead to the emergence of different physical phenomena, such as turbulence or heat transfer. Consequently, the dynamics of the entire multi-particle system could change. How the difference between the judges of entropy stability conditions affects the system must be investigated in future studies.
}
\begin{figure}[t]
\centerline{\includegraphics[scale=0.38]{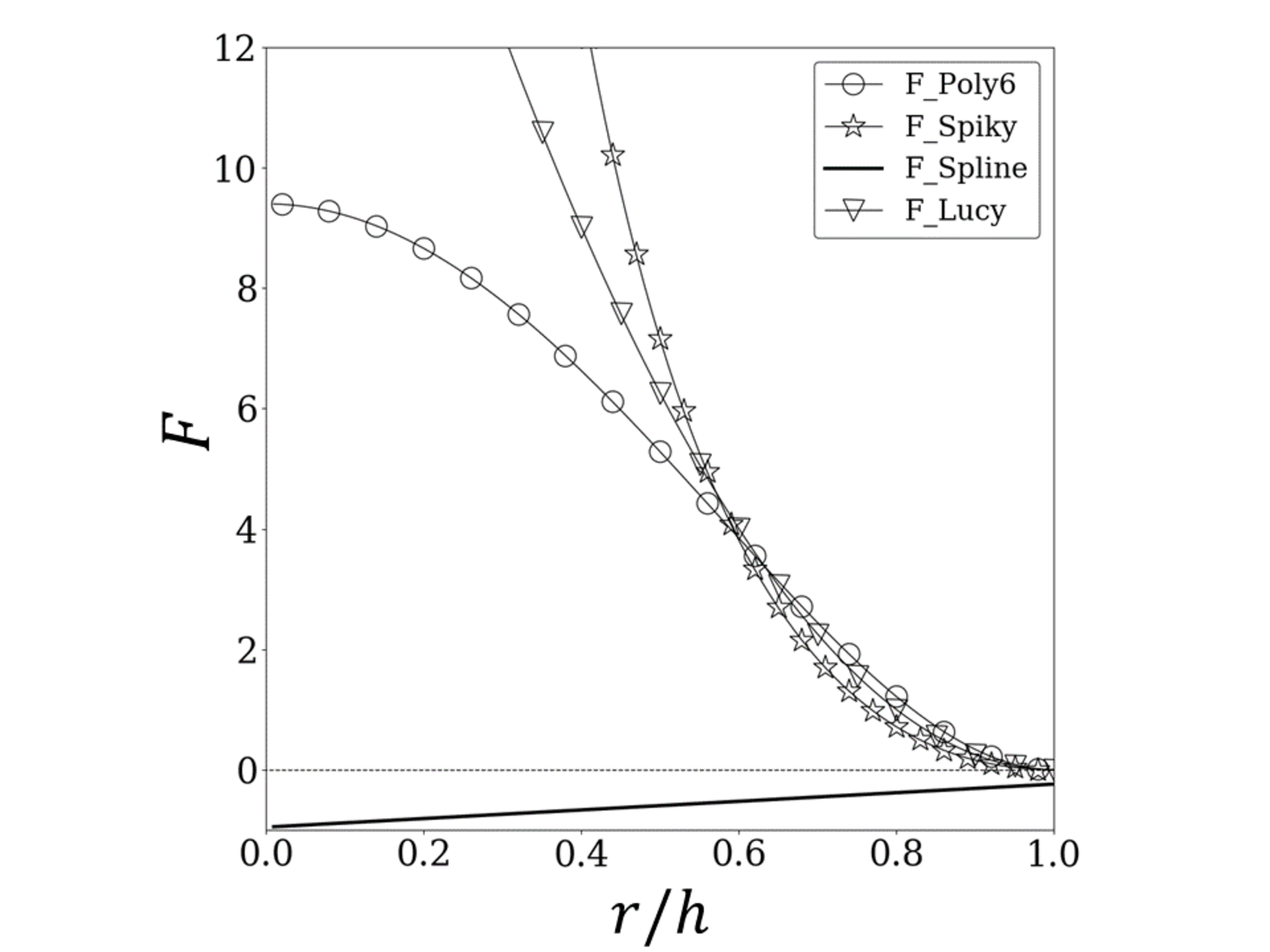}}
\HLL{\caption{Comparison of function $F$ in different types of kernel functions.}}
\label{fig:FdifferentK}
\end{figure}

\section{Conclusion}
In this article, we performed an entropy stability analysis of SDPD to evaluate the particle discretization of entropy equations exhibited in SDPD. First, \HL{we focused on the simplest SDPD system: a simulation of incompressible flows using an explicit time integration scheme, assuming a quasi-static scenario with constant volume, constant number of particles, and infinitesimal time shift.} \HL{Next,} we derived a form of entropy from the discretized entropy equation of SDPD by integrating with respect to time. Then, we examined the properties of a two-particle system for a constant temperature gradient. 

\HL{Our theoretical analysis suggests that there exist eight different types of entropy stability conditions, which depend on the types of kernel functions. It was found that the Lucy kernel, poly6 kernel, and spiky kernel produce the same types of entropy stability conditions, whereas the spline kernel produces different types of entropy stability conditions. To summarize, our results suggest that computational parameters of kernel functions contribute to the physical conditions of particle discretization systems. }

It is meaningful that we theoretically analyse the two-particle system because such a system with a small number of particles cannot be simulated owing to the lack of numerical accuracy. In that sense, our analysis in this study can be regarded as a `thought experiment'. 
Our results contribute to a deeper understanding of \HL{particle discretization}.

\section*{Acknowledgement}
This research was supported by MEXT as part of the Post-K Computer Exploratory Challenges (Exploratory Challenge 2: Construction of Models for Interaction Among Multiple Socioeconomic Phenomena, Model Development and its Applications for Enabling Robust and Optimized Social Transportation Systems; Project ID: hp190163), and partially supported by JSPS KAKENHI Grant Numbers 25287026, 15K17583, and 18H06459.
 I would also like to express my gratitude to my family for their moral support and warm encouragement.

\appendix
\section{Further analysis on a particular point of \HL{the two-particle case in the SDPD system using the Lucy kernel}}
\HL{Here, we introduce} a case where the coefficients of the second and third terms of Eq.~(\ref{eq:entropy:n2:second}) become 
zero:
\begin{eqnarray}
\phi+\alpha\left\{{\left(T_{2}^{\tau}\right)}^2-{\left(T_{2}^{0}\right)}^2 \right\} &=& 0\label{eq:t2taueqn}, \\
\phi+\alpha\left\{{\left(T_{1}^{\tau}\right)}^2-{\left(T_{1}^{0}\right)}^2 \right\} &=& 0\label{eq:t1taueqn}.
\end{eqnarray}
Equation~(\ref{eq:t2taueqn}) and Eq.~(\ref{eq:t1taueqn}) directly lead to
\begin{eqnarray}
T_{1}^{\tau} &=& \sqrt{{\Bigl(T_{1}^{0}\Bigr)}^2-\frac{\phi}{\alpha}}\label{eq:t1tauans}, \\
T_{2}^{\tau} &=& \sqrt{{\Bigl(T_{2}^{0}\Bigr)}^2-\frac{\phi}{\alpha}}\label{eq:t2tauans}.
\end{eqnarray}
In this case, we can rewrite Eq.~(\ref{eq:entropy:n2:second}) as
\begin{eqnarray}
S(\tau) & =& S(t_{0}) - 4\alpha\beta\gamma \label{eq:entropy:n2:mes},\\ 
\alpha  &:=& \frac{\bar{\kappa} F}{d^{2}}\label{eq:alpha:re},\\
\beta   &:=& \sqrt{{\Bigl(T_{1}^{0}\Bigr)}^2-\frac{\phi}{\alpha}} - {T_{1}^{0}} \label{eq:beta:n2:calc1}, \\
\gamma  &:=& \sqrt{{\Bigl(T_{2}^{0}\Bigr)}^2-\frac{\phi}{\alpha}} - {T_{2}^{0}} \label{eq:gamma:n2:calc1}.
\end{eqnarray}
Note that Eq.~(\ref{eq:alpha:re}) is a duplicate of Eq.~(\ref{eq:alpha}).
From Eq.~(\ref{eq:entropy:n2:mes}) and \HL{the positiveness of entropy in Kelvin}, we obtain the following relationship:
\begin{eqnarray}
S(\tau) &=& S(t_{0}) - 4\alpha\beta\gamma \ge 0 \label{eq:entropy:n2:compare1}, \\
\raisebox{.2ex}{.}\raisebox{1.2ex}{.}\raisebox{.2ex}{.}
~~S(t_{0}) &\ge&  4\alpha\beta\gamma.
\label{eq:entropy:n2:calc2}
\end{eqnarray}
The right-hand side of Eq.~(\ref{eq:entropy:n2:calc2}) indicates the minimum entropy at the initial time $t_{0}$.
\begin{figure}[t]
\centerline{\includegraphics[scale=0.38]{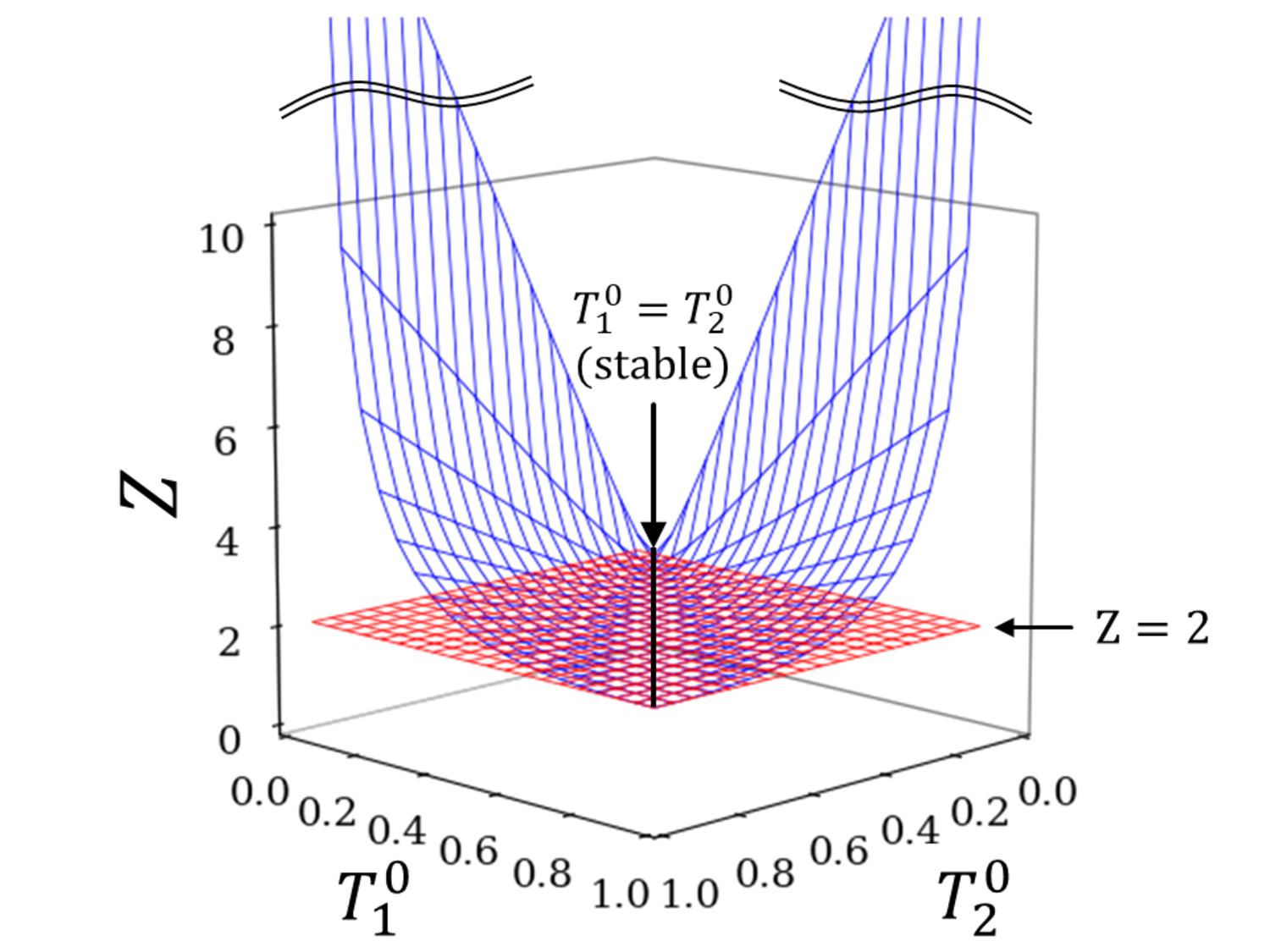}}
\caption{Stability condition of entropy $S$ at the initial time $t=t_{0}$.}
\label{fig:stabilitycond1}
\end{figure}

Here, the state of the system can be distinguished by the constant viscous heating field $\phi$ as follows:
\begin{itemize}
\item{ When $\phi \ne 0$},\\ 
the relationships $\beta\gamma > 0$ and $\alpha > 0$ are true.
Hence, the entropy of $4\alpha\beta\gamma$ on the right-hand side of \HL{Eq.~(\ref{eq:entropy:n2:calc2})} is always positive. 
\HL{There exists $\delta~( > 0)$ such that the following relationship is satisfied for all cases of $(\alpha, \beta, \gamma)$:}
\begin{eqnarray}
\exists \delta > 0,~~4\alpha\beta\gamma > \delta. \label{eq:up1}
\end{eqnarray}
Because the parameter $\alpha$ is a function of the relative position $r$ and the parameter $\beta\gamma$ is a function of the temperature $T$, 
$\alpha$ in Eq.~(\ref{eq:up1}) can be replaced with the function $f(r)$ and $\beta\gamma$ with the function $g(T)$ as
\begin{eqnarray}
\exists \delta > 0,~~f(r) g(T) > \frac{\delta}{4}. \label{eq:up2}
\end{eqnarray}

\item{When $\phi = 0$},\\
the relationship of $\beta=\gamma=0$ is true.
Hence, the value of $4\alpha\beta\gamma$ on the right-hand side of \HL{Eq.~(\ref{eq:entropy:n2:calc2})} becomes zero. Then,
\begin{eqnarray}
\phi = 0 \Longrightarrow S(\tau) = S(t_{0}) \label{eq:up3}.
\end{eqnarray}
The entropy stability condition can be simplified for $\phi = 0$ as follows: Because the relationship $T_{i}^{\tau}=T_{i}^{0}~~(i=1,2)$ is true, the first term, second term, and logarithm functions on the left-hand side of Eq.~(\ref{eq:stability:detail3}) vanish.
Because $\alpha > 0$, we obtain the stability condition as
\begin{eqnarray}
\Biggl( \frac{T_{2}^{0}}{T_{1}^{0}} + \frac{T_{1}^{0}}{T_{2}^{0}} \Biggr) \le 2. 
\label{eq:stablecond:ini}
\end{eqnarray}
Equation~(\ref{eq:stablecond:ini}) is the stability condition of the entropy $S$ at time $t_{0}$. Figure~\ref{fig:stabilitycond1} shows the \HL{plane} $Z = 2$, and the contour plot of the function $Z(T_{1}^{0}, T_{2}^{0})$ on the left-hand side of Eq.~(\ref{eq:stablecond:ini}). 
\HL{It is confirmed} that the initial state at time $t_{0}$ is stable only when $T_{1}^{0}=T_{2}^{0}$, at which point the temperature gradient becomes zero; this is consistent with fundamental statistical thermodynamics.
\end{itemize}

\bibliographystyle{h-physrev3}
\bibliography{reference}

\clearpage

\end{document}